\documentclass[epj,referee]{svjour}
\usepackage{graphicx}

\begin{document}

\title{On the Parisi-Toulouse hypothesis for the spin glass phase in 
       mean-field theory}

\author{A. Crisanti\inst{1} \thanks{andrea.crisanti@phys.uniroma1.it}
        \and T. Rizzo\inst{2} \thanks{tommaso.rizzo@inwind.it}
        \and T. Temesvari\inst{3} \thanks{temtam@helios.elte.hu}
}

\institute{ 
  \inst{1} Dipartimento di Fisica, Universit\`a di Roma 
           {\em La Sapienza}, Istituto Nazionale Fisica della 
           Materia Unit\`a di Roma I and SMC, \\
           P.le Aldo Moro 2, I-00185 Roma, Italy.\\ 
  \inst{2} SMC-INFM,  Dipartimento di Fisica, Universit\`a di Roma
           {\em La Sapienza},
           P.le Aldo Moro 2, I-00185 Roma, Italy.\\
  \inst{3} HAS Research Group for Theoretical Physics,
           E\"otv\"os University, P\'azm\'any P\'eter s\'et\'any 1/A, 
           H-1117 Budapest, Hungary.
}

\date{Received: date / Revised version: date}

\abstract{
We consider the spin-glass phase of the Sherrington-Kirkpatrick
model in the presence of a magnetic field. The series expansion of the
Parisi function $q(x)$ is computed at high orders in powers of
$\tau=T_c-T$ and $H$. We find that none  of the Parisi-Toulouse
scaling hypotheses on the $q(x)$ behavior strictly holds, although
some of them are violated only at high orders. The series is
resummed yielding results in the whole spin-glass phase which are
compared with those from a numerical evaluation of the $q(x)$. At
the high order considered, the transition turns out to be third
order on the Almeida-Thouless line, a result which is confirmed
rigorously computing the expansion of the solution near the line
at finite $\tau$. The transition becomes smoother for
infinitesimally small field while it is third order at strictly
zero field.
}


\maketitle

Mean-field theory is commonly used as a starting point
(i.e.\ zeroth order approximation) in the perturbative
treatment of finite dimensional,
short-ranged systems. As such, it is usually
very simple, almost trivial, in ordinary systems like the ferromagnet.
On the contrary, the glassy phase of disordered systems is highly
non-trivial
even in mean-field theory: in the Ising spin glass,
the prototype of such systems, the thermodynamics of the low
temperature phase is built on an ultrametrically structured order
parameter matrix proposed by Parisi (see \cite{MPV} and some of
the reprints from this book). Although the region close to the zero magnetic
field critical point is easily accessible, to get quantitative results
for physical observables deep inside the glassy phase is a rather difficult
task.

The survival of the mean-field picture, at least qualitatively,
in finite dimensional, short-ranged systems has been a controversial and
long debated problem in the last decades. It is clear that a deeper,
and more precise, understanding of the mean-field thermodynamics of
the spin glass phase is a prerequisite to make any comparison, numerical
or analytical, with physical systems. The Parisi-Toulouse (PaT) hypotheses of
Ref. \cite{PAT} (also known as projection hypotheses)
  were  not only a first step in this direction,
but had the advantage of simplicity, a common feature of
mean-field theory. An extension to systems with an average
ferromagnetic interaction followed soon \cite{T}, whereas a
scaling assumption for the order parameter function $q(x)$ was put
forward in \cite{VTG}, see eqs.\ \ref{patscal}. It was first
pointed out in ref.\ \cite{VTG} that the whole set of hypotheses
cannot be exact, as it leads to inconsistency. An argument was
presented in ref.\ \cite{TGLV} that the projection hypothesis for
the magnetization contradicts the maximization rule for the free
energy, at least near the instability, Almeida-Thouless (AT), line
\cite{AT}. An analysis of the exact, diffusion-like differential
equations describing the thermodynamics of the mean-field glassy
phase was presented in \cite{temtam}. Using series expansions,
with relatively short series, it was concluded in this paper that
the order parameter function $q(x)$ and the breakpoint $x_1$
depend only on temperature, and, among the set of statements of
the PaT hypotheses, these are the only ones which may be exact.
The order of the transition when crossing the AT-line can also be
deduced from the projection hypothesis, and it turns out to be
second order in the Ehrenfest sense, at least in finite field
\cite{PAT}. (The magnetic susceptibility has, for example, a
finite, albeit small jump.) Nevertheless, subsequent calculations
and arguments \cite{TGLV,Sommers} seemed to provide evidence that
the transition is third order, just as in zero magnetic field.
Recent numerical simulations on the SK model in a field are
discussed in \cite{BARBILL}.

In this work we study the Parisi solution of the
Sherrington-Kirkpatrick (SK) model in the presence of an external field
through the methods developed in \cite{noi}. In
particular, the computation of the high order expansion of $q(x)$
in power of the reduced temperature and of the field allows us to
finally answer the problem of the validity of the
PaT hypotheses \cite{PAT,T,VTG}. The PaT hypotheses are
a set of scaling relations concerning the behavior of the function
$q(x)$  under changes of the external fields. As we shall briefly
recall below, they  allow to compute all the quantities of interest
of the spin glass phase without solving the Parisi equations but
simply using the SK replica symmetric solution evaluated at the
AT line \cite{AT}. It turned out from our analysis
that no one of the PaT statements strictly holds, some
of them, however, are so highly consistent with the numerical estimates of
the $q(x)$ that their approximate nature can be detected only at
high order in the series expansion.

If one evaluates numerically the $q(x)$, one finds that the
magnetization seems to be independent of the temperature in the
spin glass phase; similarly the self-overlap $q_{EA}$ seems to be
independent of the magnetic field. This motivated the PaT
hypotheses which assume this fact to be exactly true, i.e.
\begin{eqnarray}
M(H,T)=M(H),  &\quad  & q_{EA}(T,H)=q_{EA}(T).
\end{eqnarray}
As a consequence, the free energy is additive, i.e.\ we have
$F(T,H)=F_1(T)+F_2(H)$. Furthermore, the energy is also additive,
while the entropy depends only on the temperature. The previous
equations show that if we want to compute $M(H)$, we can compute
the magnetization at any temperature in the spin glass phase
provided the field is equal to $H$; in particular, it can be computed
at the boundary of the spin glass phase, i.e.\ on
the AT line, where the replica symmetric solution is the correct
one:
\begin{equation}
M(H)=M(H,T)=M_{SK}(H,T_{AT}(H)).
\label{MPaT}
\end{equation}
Here the function $H_{AT}(T)$, or its inverse $T_{AT}(H)$,
parameterizes the AT line. The argument can be generalized to any
quantity that depends on either the temperature, or the magnetic
field alone. For instance, we have
\begin{equation}
S(T)=S(T,H)=S_{SK}(T,H_{AT}(T)),
\label{SPaT}
\end{equation}
\begin{equation}
q_{EA}(T)=q_{EA}(T,H)=q_{SK}(T,H_{AT}(T)).
\end{equation}
Furthermore, if we consider the expression for the energy
\begin{equation}
E=-\frac{1}{2T}\left(1-\int_0^1q^2(x)dx\right)-M H,
\end{equation}
we see that a possible way to have an additive character is
to guess proper scaling laws also for $q_0(H,T)$ and $q(x,H,T)$.
This is at the origin of the full PaT scaling laws:
\begin{equation}
\left\{
\begin{array}{lcll}
q_0(H,T)&= & q_0(H)= & q_{SK}(H,T_{AT}(H)),
\\
q(x,T,H) &= & q(x,T)= & q_u(x/T),
\\
q_{EA}(T,H)& = & q_{EA}(T). &
\end{array}
\right.
\label{patscal}
\end{equation}
The
universal temperature independent function $q_u(y)$ can be
computed
% starting
from the relation $\int_0^1q(x)dx=1-T$, which
holds in zero magnetic field, and the knowledge of $q_{EA}(T)$.
Therefore any quantity of interest can be determined simply
projecting it from the AT line, where the SK solution is correct,
down into the replica symmetry broken (RSB) phase; for this reason
the PaT hypotheses are also known as ``projection hypotheses''. In
\cite{VTG} it was already noticed that the PaT hypotheses must be
wrong somehow, this was deduced indirectly comparing the values of
different expressions for the specific heat. Instead, here we are able
to check the relations (\ref{patscal}) directly.

The equations determining $q(x)$ follows from Parisi differential
equation \cite{Parisi2} and the stationarity of the free energy. They
have appeared many times in the literature, see
ref. \cite{SDJPC84,Sommers2,temtam,noi}, therefore we choose don't
report them here. Besides the function $q(x)$ they involve two
auxiliary functions $P(x,y)$ and $m(x,y)$. Introduced as mathematical
tools to perform the computation, these functions have also a physical
interpretation: the function $m(x,y)$ represents the magnetization of
a given site properly averaged (see chap. V in Ref. \cite{MPV}) on a
cluster of states with mutual overlap $q(x)$ in presence of a frozen
effective field $y$, accordingly $m(1,y)=\tanh(\beta y)$; the function
$P(x,y)$ represents the probability distribution function over the
disorder of the effective field $y$, in particular $P(1,y)$ represents
the distribution of the cavity field \cite{MPV}. These functions have
also received a dynamical interpretation in \cite{SDJPC84}.

We have solved the equations by series expansions around $H=0$ and
$T=T_c=1$.  In the following we report the power series of the
quantities of interest at lowest orders.  To make easier the
comparison between order of magnitude we write down the series
expansion in terms of the variables $\tau=1-T$ and
$p=(3/4)^{1/3}H^{2/3}$. They are intended to be of the same order of
magnitude such that the expressions $O(k)$ below corresponds to terms
of the form $\tau^i p^{k-i}$ $(i=0,\dots,k)$. Consistently we have
$q_{EA}(T,H)=\tau+O(2)$ and $q_0(H,T)=p+O(2)$.  The computation has
been carried on up to 15th order in the order parameter by methods
similar to those described in Ref.\ \cite{noi}.  The only fact to take
care is that having two variables to expand on, time and memory grow
very fast with order, therefore we reduced the problem to a
single-variable one considering temperature-dependent values of the
field, i.e.\ we first computed the expansions at 15th order in $\tau$
assuming $p=\tau/m$ for $m=2,3\dots16$, and later we reconstructed the
$p$-dependence of the various quantities. The series are supposed to
be valid for small $\tau$ and $p$; although they are non-convergent
they can be resummed through the methods discussed in Ref.  \cite{noi}
yielding precise results in the whole spin-glass phase.
\begin{eqnarray}
q_0(H,T) & =&  p\,\left( 1 + {\frac{2\,\tau}{3}} -
     {\frac{13\,{\tau^2}}{9}} + {\frac{256\,{\tau^3}}{81}}
      \right) 
\nonumber\\
 &\phantom{=}& 
  + {p^3}\,\left( -{\frac{4}{5}} + {\frac{32\,\tau}{5}}
      \right)
-{\frac{56\,{p^4}}{27}} +O(5),
\end{eqnarray}
\begin{eqnarray}
q_{EA}(T,H) &= & \tau + {\tau^2} - {\tau^3} +
  {\frac{5\,{\tau^4}}{2}} - {\frac{171\,{\tau^5}}{10}} +
  {\frac{1077\,{\tau^6}}{10}}
\nonumber\\
 &\phantom{=}& 
+  {p^5}\,\left( {\frac{8}{5}} - {\frac{208\,\tau}{15}}
      \right)
-
{\frac{16\,{p^6}}{9}} +O(7),
\end{eqnarray}

\begin{eqnarray}
q(x,T,H)&-&q(x,T,0) =
\nonumber\\
 &\phantom{=}& \phantom{+}
p^5\left({32 \over 5}-{256 \tau \over 15}+{3808 \tau^2\over 45}\right)
\nonumber\\
 &\phantom{=}& 
+ p^6\left({-64 \over 9}+{64 \tau \over 3}\right)+{2144 \over 35}p^7
\nonumber\\
 &\phantom{=}& 
+\left( {\frac{-132\,{p^5}}{5}} +
     {\frac{88\,{p^6}}{3}} + {\frac{448\,{p^5}\,\tau}{5}}
      \right) \,x 
\nonumber\\
 &\phantom{=}& 
+ {\frac{312\,{p^5}\,{x^2}}{5}}+O(8),
\end{eqnarray}
\begin{eqnarray}
f(T,H) & = &- (1-\tau)
  \ln 2-{\frac{1}{4}}-{\frac{\tau}{4}}  - {\frac{\tau^2}{4}} -
  {\frac{{\tau^3}}{12}} + {\frac{{\tau^4}}{24}} 
\nonumber\\
 &\phantom{=}& 
- {\frac{{\tau^5}}{120}} 
+ {\frac{3\,{\tau^6}}{20}} 
- {\frac{79\,{\tau^7}}{140}} 
+ {\frac{1679\,{\tau^8}}{560}}
- {\frac{2\,{p^3}}{3}} 
\nonumber\\
 &\phantom{=}& 
+ {\frac{2\,{p^5}}{5}} 
- {\frac{8\,{p^6}}{27}} 
- {\frac{11\,{p^7}}{35}} 
+ {\frac{50\,{p^8}}{27}}
\nonumber\\
 &\phantom{=}& 
+ {p^5}\,\left( {-\frac{4\,\tau}{15}} +
     {\frac{34\,{\tau^2}}{45}} - {\frac{872\,{\tau^3}}{405}}
      \right)
\nonumber\\
 &\phantom{=}& 
+
  {p^6}\,\left( {\frac{8\,\tau}{27}} -
     {\frac{8\,{\tau^2}}{9}} \right)
+
{\frac{-52\,{p^7}\,\tau}{105}}  +O(9).
\label{FPT}
\end{eqnarray}
The expansions of energy and magnetization have been computed from the
following formulas \cite{SDJPC84,temtam,noi}
\begin{eqnarray}
M &=&\int P(x,y)\, m(x,y)\, dy \\
E &=&-{\beta \over 2}\left(1-\int_0^1q^2(x)dx\right)-M H.
\end{eqnarray}
It turned out that they verify all the usual thermodynamic relations
such as $S=-\partial f/\partial T$
\footnote{In Ref.\ \cite{temtam}, it was suggested that this thermodynamic
relation may not be valid in the spin glass phase.
The argument was based on an extrapolation of relatively low
order results. It turns out from the analysis of the higher order terms that
quantities which were assumed in \cite{temtam} to depend on the
temperature alone, do have a magnetic field dependence too. These findings change
the conclusions basically, leading to a support of the validity of that
thermodynamic relation.}
and $m=-\partial f/\partial H$.

\begin{figure}[tbp]
\begin{center}
\resizebox{0.9\columnwidth}{!}{
\includegraphics{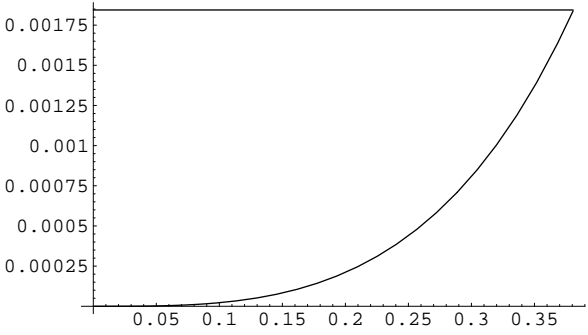}
} 
\caption{Plot of $q_{EA}(T,H)-q_{EA}(T,0)$ at
fixed $T=.6$ as a function of $H$ for $0\leq H\leq
H_{AT}(T)=.3826$, the straight line is
$q_{SK}(T,H_{AT}(T))-q_{EA}(T,0)=.50689-.50504=.00184$ at $T=.6$.
The data were obtained by  the point-by-point procedure for
multivariate expansions \protect\cite{noi} using
the $(7,7)$ Pad\'e approximant. For small $H$, the data fit a
power-law behavior $.053854\, H^{10/3}$.} 
\label{figureq1}
\end{center}
\end{figure}

\begin{figure}[tbp]
\begin{center}
\resizebox{0.9\columnwidth}{!}{
\includegraphics{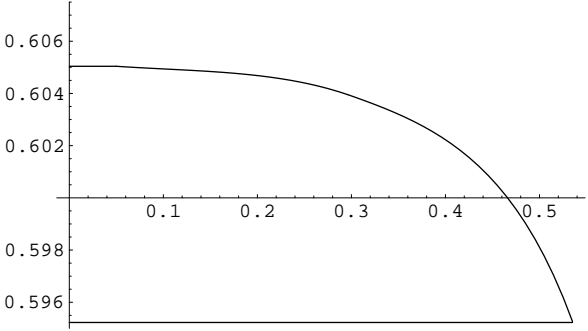} 
}
\caption{Plot of $q_0(H,T)$ at fixed $H=.5$ as
a function of $T$ for $0\leq T\leq T_{AT}(H)=.5351$, the straight
line is $q_{SK}(H,T_{AT}(H))=.5951$ at $H=.5$. The data were
obtained by  the point-by-point procedure for multivariate
expansions \protect\cite{noi} using the $(7,7)$,
$(6,8)$ and $(8,6)$ Pad\'e approximants.
}
\label{figureq0}
\end{center}
\end{figure}

>From our analysis the following conclusions can be
drawn
regarding the PaT scaling hypothesis.
\begin{itemize}

\item $q_{EA}(T,H)=q_{EA}(T)$. This scaling law is violated in the
expansion in power of $\tau$ and $p$ at order $p^5$ (i.e.\
$H^{10/3}$) which means fifth order in the height $q_0(T,H)$ of
the first plateau. As a consequence, such a violation is hardly
seen numerically, and the PaT approximation
$q_{EA}(T)=q_{SK}(T,H_{AT}(T))$ is excellent at high temperature.
We note that if this relation were true, the expansion of
$q_{EA}(T)$ around $T=0$ should read $q_{EA}(T)=1-3/2T^2+O(T^3)$.
This is the expression which is valid on the AT line while the
true expression is $q_{EA}(T)=1-\alpha T^2+O(T^3)$ with
$\alpha=1.60\pm .01$ obtained resumming the expansion in power of
$\tau$ and numerically \cite{noi}.
In figure \ref{figureq1} we plot $q_{EA}(T,H)-q_{EA}(T,0)$ at fixed $T$ as a function of the magnetic field. The data were obtained from the series expansions  by the resummation procedures discussed in Ref.\ \cite{noi}.

\item $q(x,T)=q_u(\beta x)$ is violated near the critical
temperature, as can be seen from the series expansion \cite{noi}. However, the
scaling is a very good approximation in a wide range of
temperatures. It becomes exact in the limit $T\rightarrow 0$ and
$x/T\rightarrow 0$.  A consequence of this scaling law would be
$\lim_{T\rightarrow 0}x_1=1/2$ \cite{VTG}, while the real value
turns out to be $.548 \pm .005$ \cite{noi}. From the above power series it can be also deduced that $x_1(T,H)$ picks up a small dependence on $H$, again proportional to $p^5$ (i.e.\ $H^{10/3}$).

\item  $q(x,T,H)=q(x,T)$. This scaling law is violated in the
expansion in power of $\tau$ and $p$ at order $p^5$ (i.e.
$H^{10/3}$) which means fifth order in the height $q_0(T,H)$ of
the first plateau. As for $q_{EA}(T)$, the deviation is very
difficult to be seen numerically. Notice that this scaling law is
also valid near $T_c$ when the more stringent $q(x,T)=q_u(x/T)$ is
not verified. We have checked at 15th order that all the
coefficients in the expansion of $q(x,T,H)$ in power of $x$ pick up
corrections proportional to $p$ such that the lowest power of $p$
is always $p^5$ (i.e.\ $H^{10/3}$). This is likely to remain true
at all orders.

\item  $q_0(H,T)=q(H)$. This scaling law is violated too. However,
at all values of the magnetic field the behavior of the function
$q_0(H,T)$ is similar to that of figure \ref{figureq0} obtained by
resumming the series expansion. Its value increases linearly from
the PaT value at $T=T_{AT}(H)$ while lowering the temperature, but
it goes rapidly to a constant value at low temperature.
Furthermore, at all fields the difference between the PaT value
$q_0(H,T_{AT}(H))$ and $q_0(H,0)$ is always of order $10^{-2}$.
Actually, this is the only violation of the PaT scaling which can
be easily seen numerically.

\item $F(T,H)=F_1(T)+F_2(H)$. The first term in the expansion in
power of $\tau$ and $p$ which violates this additive law is $\tau
p^5$ i.e.\ $\tau H^{10/3}$, a sixth order term. As a consequence,
the deviation from the PaT entropy (\ref{SPaT}) is only fifth
order, and again it is a very good approximation. The same is
true for the PaT magnetization (\ref{MPaT}). We have checked at
very high order in the expansion that the lowest power of $p$ in
the mixed terms is always $p^5$ (i.e.\ $H^{10/3}$), i.e.\ fifth
order in the height $q_0$ of the first plateau, a result that is
likely to be valid at all temperatures, explaining the goodness of
the PaT approximation in the whole RSB phase. Indeed, it turns out
that the approximation $q(x,T,H)=q(x,T,0)$ breaks down at fifth
order in the expansion in powers of $q_0$ at a given {\em finite}
$\tau$. Therefore we expect that at all temperatures the
corrections to the PaT estimates are of order $q_0^5$, and become
relevant only at very high fields. In particular, the field may be
expressed up to fourth order in $q_0$ in terms of the function
$q(x,T,0)$, we have indeed $h^2=2/(3\,T q'(0,T))q_0^3+0\,
q_0^4+O(q_0^5)$; this explains why the scaling $q_0(H,T)=q(H)$ is
best verified at low temperature, since $T q'(0,T)$ tends to
the constant value $0.743\pm 0.002$ for $T\rightarrow 0$
\cite{noi}.
\end{itemize}

The predictions from the perturbative expansions have been
compared  with those  from the numerical solution of the equations
for the $q(x)$ using the high precision technique introduced in
ref. \cite{noi}. By comparing the $H=0$ with the $H\neq 0$
solution  in all cases we found a rather good agreement with
analytical results and the violation of the PaT scaling can be
appreciated.

The series expansion at high orders allows us to safely determine the
order of the transition on the AT line. By making an expansion of
the replica symmetric (RS) solution near the transition line the following conclusions can be drawn.

\begin{itemize}

\item
The RS-RSB transition is third order, i.e.\ all
the possible derivatives of the free energy (temperature, field and mixed)
up to second order, inclusively, are continuous along the AT line.
All higher order derivatives are discontinuous, and not divergent, at a generic point on the AT line.

\item 
Near the critical
point ($\tau=0$, $H=0$) the situation is more complicated.
First of all, we note that the RS free energy is singular at this
point, therefore its derivatives depend on how this
point is approached. For instance, temperature
derivatives   of the RS free energy higher than the fifth are divergent approaching the
critical point on the AT line, while they are regular on the line ($H=0$, $T>1$).

\item
Approaching the critical
point ($\tau=0$, $H=0$)
along the AT line, the discontinuity  in the third field derivative goes to
zero as $\sqrt{\tau}$, while the fourth derivative of the RS and
RSB free energy (and also their difference) diverge.
%{\it Comments on the divergence of the fourth derivatives}.
This fourth derivative with respect to the magnetic field is essentially
the nonlinear susceptibility, and it can be expressed, by a
fluctuation-dissipation-like formula, in term of the so-called longitudinal
correlation function ($G_L$). $G_L$ becomes identical with the spin glass
susceptibility in exactly zero magnetic field, and diverges at $T_c$
($\tau=0$), although it remains finite elsewhere.

\item
The discontinuities in the third
and fourth temperature derivatives also go to zero while
approaching the critical point along the AT line, only the fifth temperature
derivative has a finite discontinuity in this limit.
As already mentioned, higher temperature derivatives of the RS free energy diverge
at $T_c$, while the RSB derivatives are regular.
Therefore, if we consider only temperature derivatives, the RS-RSB
transition becomes fifth order near the critical point. 
This can be understood
noticing that, while approaching the critical point from the left
along the AT line, the RS solution, valid above the line, is similar to  
the (unstable) RS solution at $T<1$ and $H=0$ whose free energy
differs from the RSB free energy at fifth order in $\tau$.
This behaviour of the temperature derivatives on the AT line in the  $H\rightarrow 0$ limit is markedly different from that at $H$ 
strictly zero. Indeed at $H=0$ we must consider the $T>1$ RS solution whose free energy
differs from the RSB free energy at third order in $\tau$, as can be checked using $F_{RS}=-1/(4T)-T\log 2$ valid for $T>1$.

\end{itemize}

These results were obtained analyzing very long series and are
rather safe, however to confirm them we performed a series
expansion of the RSB solution near the AT line in power of
$dh=h-h_{AT}$ at fixed temperature. This is an improvement on the
result of \cite{Sommers} where such an expansion was obtained
using a 1RSB function instead of the correct full RSB $q(x)$. This
expansion allows to compute exactly the derivatives of the RSB
free energy on the AT line. As expected we have found that the
first and second field derivatives of the RS and RSB free energies
are continuous on the AT line. In general we have observed  that
in order to determine the free energy at order $dh^n$ it is
sufficient to know $q(x)$ at order $dh^{n-2}$ instead of order
$dh^{n-1}$ which would be expected by power counting. Therefore
the RS solution, which is the zeroth order approximation, gives
the correct free energy up to second order in $dh$. The
coefficients of this expansions depend on integrals of the form
\begin{equation}
\overline{m^j}={1 \over \sqrt{2 \pi q}}\int_{-\infty}^{+\infty}e^{-y^2/(2q)}\tanh[\beta y+\beta H]^j dy
\end{equation}
evaluated on the AT line. For instance the second field derivative reads
\begin{eqnarray}
{\partial^2 F \over \partial H^2}(&T&,H_{AT}(T)) =
   \frac{\beta}{1+(q_{AT}-1)\beta^2} 
\nonumber\\
&\phantom{=}& 
\times \Bigl(q_{AT}-1-2 q_{AT}\beta^2+q_{AT}^2\beta^2
\nonumber\\
&\phantom{}& 
-\beta^2\bigl(-1+M_{AT}^2-2M_{AT}\overline{m^3}+\overline{m^3}^2\bigr)\Bigr)
\end{eqnarray}
The breakpoint reads
\begin{equation}
x_1(T,H_{AT}(T))=\frac{2(-2+2 \beta^2-3q_{AT}\beta^2+\beta^2 \overline{m^6})}{3-2 \beta^2+3q_{AT}\beta^2-\beta^2 \overline{m^6}}
\end{equation}
It is interesting to notice that the $T\rightarrow 0$ limit of the
breakpoint is exactly $1/2$ as predicted by the PaT hypotheses
in an independent way and at variance with the $(T=0,H=0)$ value
which is $.548\pm .005$ \cite{noi}.

\section*{Acknowledgments}
One of us (T.\ T.) acknowledges support from the Hungarian Science
Fund (OTKA), Grant No.~T032424.


\begin{thebibliography}{99}

\bibitem{MPV}
        M.~M\'ezard, G.~Parisi, and M.A.~Virasoro,
        "Spin glass theory and beyond", World Scientific (Singapore 1987).

\bibitem{Parisi2}
        G. Parisi, 
	J. Phys. A {\bf 13}, (1980) L115.

\bibitem{Parisi3}
        G. Parisi, 
	J. Phys. A  {\bf 13}, (1980) 1887.

\bibitem{PAT}  
  G. Parisi and G. Toulouse, 
  J. Physique Lett. {\bf 41},  (1980) L361.

\bibitem{T}
  G. Toulouse, 
  J. Physique Lett. {\bf 41}, (1980) L447.

\bibitem{VTG}
  J. Vannimenus, G. Toulouse and G. Parisi,
  J. Physique {\bf 42}, (1981) 565.

\bibitem{TGLV}
  G. Toulouse, M. Gabay, T.C. Lubensky and J. Vannimenus,
  J. Physique Lett. {\bf 43}, (1982) L109.

\bibitem{AT}
        J. R. L. de Almeida and D. J. Thouless, 
	J. Phys. A {\bf 11}, (1978) 983.

\bibitem{temtam}
  T. Temesv\'ari, J. Phys. A {\bf 22}, (1989) L1025.

\bibitem{Sommers}
  H.-J. Sommers, 
  J. Phys. A {\bf 17}, (1984) 2351.

\bibitem{Sommers2} 
  H.-J. Sommers, J. Physique Lett. {\bf 46}, (1985) L779.

\bibitem{noi}
        A. Crisanti and T. Rizzo, 
	Phys. Rev. E {\bf 65}, (2002) 046137.

\bibitem{SDJPC84} 
  H.-J. Sommers, W. Dupont, 
  J. Phys. C {\bf 17}, (1984) 5785.

\bibitem{BARBILL}
        A. Billoire and B. Coluzzi, 
        Phys. Rev. E {\bf 67}, (2003) 036108.

\end{thebibliography}
\end{document}